\documentclass[twocolumn]{aastex631}

\newcommand{\qfit}{\texttt{q3dfit}}
\newcommand{\pyqsofit}{\texttt{pyqsofit}}

\newcommand{\oi}{\hbox{[O$\,${\scriptsize I}]}}

\newcommand{\oiii}{\hbox{[O$\,${\scriptsize III}]}}

\newcommand{\nii}{\hbox{[N$\,${\scriptsize II}]}}
\newcommand{\sii}{\hbox{[S$\,${\scriptsize II}]}}
\newcommand{\ha}{\hbox{H$\alpha$}}
\newcommand{\hb}{\hbox{H$\beta$}}

\newcommand{\siiha}{log([S{\sc II}]/\ha)}
\newcommand{\oiha}{log([O{\sc I}]/\ha)}

\newcommand{\kms}{km\,s$^{-1}$} 
\newcommand{\msun}{M$_{\odot}$} 
\newcommand{\eden}{cm$^{-3}$} 

\newcommand{\momfluxagn}{$\dot{P}_{QSO}$ }

\newcommand{\momfluxratio}{$\frac{\dot{P}_{outflow}}{\dot{P}_{AGN}}$}

\newcommand{\ergs}{erg s$^{-1}$ }
\newcommand{\myr}{M$_\odot$~yr$^{-1}$}

\renewcommand{\deg}{\ensuremath{^{\circ}}}
\newcommand{\ferg}{erg s$^{-1}$ cm$^{-2}$ }
\newcommand{\surff}{erg s$^{-1}$ cm$^{-2}$ \AA$^{-1}$arcsec$^{-2}$}
\newcommand{\htwo}{H$_{2}$}

\shorttitle{\textit{JWST} observations of quasar-driven outflow}
\shortauthors{Vayner et al.}
\graphicspath{{./}{figures/}}

\begin{document}

\title{First results from the JWST Early Release Science Program Q3D: Powerful quasar-driven galactic scale outflow at $z=3$}

\correspondingauthor{Andrey Vayner}
\email{avayner1@jhu.edu}

\author[0000-0002-0710-3729]{Andrey Vayner}
\affiliation{Department of Physics and Astronomy, Bloomberg Center, Johns Hopkins University, 3400 N. Charles St., Baltimore, MD 21218, USA}

\author[0000-0001-6100-6869]{Nadia L. Zakamska}
\affiliation{Department of Physics and Astronomy, Bloomberg Center, Johns Hopkins University, 3400 N. Charles St., Baltimore, MD 21218, USA}

\author[0000-0001-7572-5231]{Yuzo Ishikawa}
\affiliation{Department of Physics and Astronomy, Bloomberg Center, Johns Hopkins University, 3400 N. Charles St., Baltimore, MD 21218, USA}

\author[0000-0002-4419-8325]{Swetha Sankar}
\affiliation{Department of Physics and Astronomy, Bloomberg Center, Johns Hopkins University, 3400 N. Charles St., Baltimore, MD 21218, USA}

\author[0000-0003-2212-6045]{Dominika Wylezalek}
\affiliation{Zentrum für Astronomie der Universität Heidelberg, Astronomisches Rechen-Institut, Mönchhofstr 12-14, D-69120 Heidelberg, Germany}

\author[0000-0002-1608-7564]{David S. N. Rupke}
\affiliation{Department of Physics, Rhodes College, Memphis, TN 38112, USA}

\author[0000-0002-3158-6820]{Sylvain Veilleux}
\affiliation{Department of Astronomy and Joint Space-Science Institute, University of Maryland, College Park, MD 20742, USA}

\author[0000-0002-6948-1485]{Caroline Bertemes}
\affiliation{Zentrum für Astronomie der Universität Heidelberg, Astronomisches Rechen-Institut, Mönchhofstr 12-14, D-69120 Heidelberg, Germany}

\author[0000-0003-2405-7258]{Jorge K. Barrera-Ballesteros}
\affiliation{Instituto de Astronomía, Universidad Nacional Autónoma de México, AP 70-264, CDMX 04510, Mexico}

\author[0000-0001-8813-4182]{Hsiao-Wen Chen}
\affiliation{Department of Astronomy \& Astrophysics, The University of Chicago, 5640 South Ellis Avenue, Chicago, IL 60637, USA}

\author{Nadiia Diachenko}
\affiliation{Department of Physics and Astronomy, Bloomberg Center, Johns Hopkins University, 3400 N. Charles St., Baltimore, MD 21218, USA}

\author[0000-0003-4700-663X]{Andy D. Goulding}
\affiliation{Department of Astrophysical Sciences, Princeton University, 4 Ivy Lane, Princeton, NJ 08544, USA}

\author[0000-0002-5612-3427]{Jenny E. Greene}
\affiliation{Department of Astrophysical Sciences, Princeton University, 4 Ivy Lane, Princeton, NJ 08544, USA}

\author[0000-0003-4565-8239]{Kevin N. Hainline}
\affiliation{Steward Observatory, University of Arizona, 933 North Cherry Avenue, Tucson, AZ 85721, USA}

\author{Fred Hamann}
\affiliation{Department of Physics \& Astronomy, University of California, Riverside, CA 92521, USA}

\author[0000-0001-8813-4182]{Timothy Heckman}
\affiliation{Department of Physics and Astronomy, Bloomberg Center, Johns Hopkins University, Baltimore, MD 21218, USA}

\author[0000-0001-9487-8583]{Sean D. Johnson}
\affiliation{Department of Astronomy, University of Michigan, Ann Arbor, MI 48109, USA}

\author{Hui Xian Grace Lim}
\affiliation{Department of Physics, Rhodes College, Memphis, TN 38112, USA}

\author[0000-0003-3762-7344]{Weizhe Liu}
\affiliation{Department of Astronomy, Steward Observatory, University of Arizona, Tucson, AZ 85719, USA}

\author[0000-0003-0291-9582]{Dieter Lutz}
\affiliation{Max-Planck-Institut für Extraterrestrische Physik, Giessenbachstrasse 1, D-85748 Garching, Germany}

\author[0000-0001-6126-5238]{Nora Lützgendorf}
\affiliation{European Space Agency, Space Telescope Science Institute, Baltimore, Maryland, USA}

\author[0000-0002-1047-9583]{Vincenzo Mainieri}
\affiliation{European Southern Observatory, Karl-Schwarzschild-Straße 2, D-85748 Garching bei München, Germany}

\author{Ryan McCrory}
\affiliation{Department of Physics, Rhodes College, Memphis, TN 38112, USA}

\author[0009-0007-7266-8914]{Grey Murphree}
\affiliation{Department of Physics, Rhodes College, Memphis, TN 38112, USA}
\affiliation{Institute for Astronomy, University of Hawai'i, 2680 Woodlawn Dr., Honolulu, HI, 96822, USA}

\author[0000-0001-5783-6544]{Nicole P. H. Nesvadba}
\affiliation{Université de la Côte d'Azur, Observatoire de la Côte d'Azur, CNRS, Laboratoire Lagrange, Bd de l'Observatoire, CS 34229, Nice cedex 4 F-06304, France}

\author[0000-0002-3471-981X]{Patrick Ogle}
\affiliation{Space Telescope Science Institute, 3700, San Martin Drive, Baltimore, MD 21218, USA}

\author[0000-0002-0018-3666]{Eckhard Sturm}
\affiliation{Max-Planck-Institut für Extraterrestrische Physik, Giessenbachstrasse 1, D-85748 Garching, Germany}

\author{Lillian Whitesell}
\affiliation{Department of Physics, Rhodes College, Memphis, TN 38112, USA}

\begin{abstract}
Quasar-driven galactic outflows are a major driver of the evolution of massive galaxies. We report observations of a powerful galactic-scale outflow in a $z=3$ extremely red, intrinsically luminous ($L_{\rm bol}\simeq 5\times 10^{47}$\ergs) quasar SDSSJ1652+1728 with the Near Infrared Spectrograph (NIRSpec) on board JWST. We analyze the kinematics of rest-frame optical emission lines and identify the quasar-driven outflow extending out to $\sim 10$ kpc from the quasar with a velocity offset of  ($v_{r}=\pm 500$ \kms) and high velocity dispersion (FWHM$=700-2400$ \kms). Due to JWST's unprecedented surface brightness sensitivity in the near-infrared- we unambiguously show that the powerful high velocity outflow in an extremely red quasar (ERQ) encompasses a large swath of the host galaxy's interstellar medium (ISM). Using the kinematics and dynamics of optical emission lines, we estimate the mass outflow rate -- in the warm ionized phase alone -- to be at least $2300\pm1400$ $M_{\odot}$ yr$^{-1}$. We measure a momentum flux ratio between the outflow and the quasar accretion disk of $\sim$1 on kpc scale, indicating that the outflow was likely driven in a relatively high ($>10^{23}$cm$^{-2}$) column density environment through radiation pressure on dust grains. We find a coupling efficiency between the bolometric luminosity of the quasar and the outflow of 0.1$\%$, matching the theoretical prediction of the minimum coupling efficiency necessary for negative quasar feedback. The outflow has sufficient energetics to drive the observed turbulence seen in shocked regions of the quasar host galaxy, likely directly responsible for prolonging the time it takes for gas to cool efficiently.

\end{abstract}

\keywords{}

\section{Introduction} \label{sec:intro}

Quasar feedback is the interaction between the energy output of an accreting supermassive black hole (SMBH) and its surrounding environments that may be an integral part of galaxy evolution. Quasar feedback is often invoked to explain the quenching of star formation, regulating black hole mass, and limiting galaxy growth \citep{silk98,crot06,fabi12,korm13}, especially in the most massive ($>10^{12.5}$\msun) dark matter halos in the local Universe. Empirical evidence of tight correlations between the SMBH and the host galaxy properties \citep{mago98,gebh00} suggests that there may be a strong physical connection, perhaps a co-evolution, between SMBHs and galaxies. 

Quasar feedback often manifests itself as either radiation pressure driven winds \citep{murr95} or jets \citep{wagn12} that interact with the immediate interstellar medium and drive powerful multi-gas-phase outflows that propagate into the galaxy and expel gas, drive shocks, and prolong the time it takes for gas to cool and form stars. Only a small fraction of the quasar bolometric luminosity needs to be converted into the kinetic luminosity of the galaxy scale outflow for negative feedback to play an important role in regulating star formation \citep{choi12,hopk12}. While generally, quasar feedback is expected to have an overall negative impact on the long time scale, outflows can, in specific situations, compress gas ahead of the flow that can momentarily trigger star formation, causing temporary positive feedback \citep{Silk13,Maiolino17,Mercedes-Feliz23}. 

At low redshifts, much of the stellar components of massive galaxies are already set, suggesting that quasar feedback must have had a greater impact at earlier epochs when galaxies and black holes were experiencing more rapid growth. Both galaxy growth and quasar activity peak at cosmic noon around $z\sim1-3$ \citep{mada14}, so it is likely that quasar feedback may be most important at this epoch \citep{boyl98,hopk08}. A large number of studies reveal high redshift quasars with forbidden emission lines with large velocity offsets and/or dispersions indicative of quasar-driven winds. The advent of IFU spectroscopy, especially with adaptive optics from the ground on 8-10m class telescopes, reveal that a large number of these outflows are on galactic scales with energetics powerful enough to affect the star formation properties of their host galaxies \citep{nesv08,cano12,Vayner17,Vayner21c,Kakkad20}. In a select few sources, multi-gas phase outflows detected in both warm ionized and molecular phases indicate direct evidence for negative quasar feedback due to the removal of the molecular reservoir, with the outflow often being the dominant source of gas depletion and a substantial fraction of the galaxies' molecular gas in an outflow \citep{Brusa18,Herrera-Camus19,Vayner21c}.   

The onset of quasar feedback is likely to be most efficient in obscured nuclear environments where it is easier to couple the energy and momentum of the photons from the quasar accretion disk to the immediate ISM \citep{fauc12b}. Quasars accreting at high Eddington ratios in obscured environments are likely to be ideal candidates to study the early phases of quasar feedback and catch the transitional phase from an obscured to the unobscured galactic nucleus phase, often referred to as the ``blow-out" phase. 

Dust-obscured active galactic nuclei (AGN) are thus the preferred population to study this short but essential epoch of feedback. Specifically, high-redshift radio galaxies (HzRGs), hot dust-obscured galaxies (HotDOGs), and extremely red quasars (ERQs) are three key samples of obscured powerful AGN in the distant Universe to study the feedback phenomena and understand the short-lived, high-impact phase of quasar feedback. All three samples show strong evidence for powerful AGN-driven outflows \citep{nesv17,perr19,Finnerty20,Vayner21a}. 

Of these obscured populations, ERQs may exhibit one of the most extreme form of quasar feedback. ERQs have been known to show extended, fast-moving, powerful outflows, as traced by the ionized \oiii\ gas, and are believed to be near/super-Eddington accretors \citep[e.g.,][]{zaka16b,alex18,perr19,Vayner21a}. ERQs reside in the region of the column density vs. Eddington ratio diagram where quasars are predicted to undergo a fast blowout phase, with nuclear column densities $>10^{23}$cm$^{-2}$ and Eddington ratios $>1$ \citep{Ishibashi18,Lansbury20,Ishikawa21}. These outflows may be signposts of quasar feedback's ``blow-out' phase. Thus, studying the feedback in ERQs may provide insights into the impact the outflows may have on the host galaxy. 

Dust reddening combined with cosmic surface brightness dimming makes studying obscured quasars and their host galaxies difficult. Detailed studies at high redshift have been challenging until \textit{JWST}, which enables, for the first time, a detailed look at the ionized outflows from a $z\sim3$ quasars and the faint emission from the host galaxy. 

This paper follows up on our results from \citep{Wylezalek22,Vayner23b} that focused on the immediate quasar host galaxy environment and the ionization properties of the gas. This paper presents a detailed emission-line study of the warm ionized outflow in SDSSJ1652+1728 (hereafter J1652). We aim to understand the extent and morphology of the outflow, the energetics, the dominant driving mechanism behind the galaxy scale outflow, and the impact on the quasar host galaxy. We summarize the observation and data reduction in Sections \ref{sec:obs_data_reduction}, we explore the multi-wavelength properties in Section \ref{sec:multi-wave}, we present the NIRSpec data analysis, emission line fitting and outline the different kinematic components in Section \ref{sec:analysis}. We estimate the energetics in Section \ref{sec:out-energy}, we discuss the driving mechanism behind the outflow in Section \ref{sec:discussion} and compare it to other ionized outflows in powerful quasars. Finally, we summarize our results in Section \ref{sec:conclusions}

\section{Observations and Data reduction} 
\label{sec:obs_data_reduction}

In this section, we outline the observations, data reduction, and analysis of the \textit{JWST} NIRSpec data.

\subsection{Observations}

Observations of J1652 were taken on 2022-07-15 beginning at 22:41:17.593 UTC. We used the G235H grating in combination with the F170LP filter. The source was acquired using the ``point and shoot" method. Observations were taken using the ``NRSIRS2" readout mode with 25 non-destructive group reads where 5 frames were averaged to combine into a single group corresponding to a total exposure time per integration of 1823.6s. The object was dithered in the 3\arcsec $\times$3\arcsec\ NIRSpec field of view using a 9-point small cycling pattern to improve the spatial sampling of the PSF. The total on-source exposure time was 16412.5 s. We took a single exposure with all the MSA shutters and IFU aperture closed to remove any light leakage from bright objects in the NIRSpec instrument field of view and to remove any background light from failed open shutters. The leak calibration exposure was taken at the first science dither position using the same detector readout and exposure time. 

\section{NIRSpec Data reduction data Reduction} \label{sec:data-reduction}

For data reduction, we use the Space Telescope (ST) \textit{JWST} pipeline version 1.8.2. The first stage of the pipeline, ``Detector1Pipeline" performs standard infrared detector reduction steps such as dark current subtraction, fitting ramps of non-destructive group readouts, combining groups and integrations, data quality flagging, cosmic ray removal, bias subtraction, linearity and persistence correction. \\

Afterward, we run ``Spec2Pipeline" which assigns a world coordinate system to each frame, flat field correction, flagging pixels affected by open MSA shutters, and extracts the 2D spectra into a 3D data cube using the ``cube build" routine. We select the ``emsm" method to extract the 2D spectra into a 3D data cube using 100 mas plate scale. We skip the imprint subtraction step since it increases the final data cube's overall noise and creates stronger background variations at specific wavelength channels. \\

To combine the different dither positions, we use an in-house script based on the Python \texttt{Reproject} package \citep{thomas_robitaille_2023_7584411}, which finds a common WCS system between the dithered observations and aligns the data to match spatially reprojecting each dithered cube onto a 50 mas plate scale with the flux-conserving routine. Before combining the individual data cubes, we run a sigma clip routine that determines outliers across the 9-dither position. We mask the outliers and exclude them from the data sets. We average together the 9 data cubes into a single frame. We perform the same data reduction for both NRS detectors. We stitch the two data cubes from each detector together by interpolating the spectral axis of both data sets onto a common spectral grid. A similar process is done on the variance data cubes produced by the ``Spec2Pipeline". \\

Similar data analyses are done on the standard star TYC 4433-1800-1 (proposal ID: 1128). We extract the spectrum of the standard star over a large radial aperture. We extract a model near-infrared spectrum of this star from the HST standard star catalog. We scale the model spectrum to match the UV and optical flux from STIS observations. We interpolate the near-infrared model onto the wavelength range of the G235H grating observations using linear interpolation. We then divide the standard star spectrum by the model to create a conversion array into cgs flux units. We multiply our science observations by the flux conversion array. We achieve a final 2$\sigma$ flux sensitivity of 2.87$\times10^{-19}$ \surff\ and AB magnitude/arcsec$^{2}$ of 22.45 at 1.995 \micron, near redshifted \oiii\ 5007 \AA\ and 2.45$\times10^{-19}$ \surff\ and AB magnitude/arcsec$^{2}$ of 22.09 at 2.549 \micron, near redshifted \ha\ in a 0.4\arcsec$\times$0.4\arcsec aperture.

\section{Multi-wavelength data} \label{sec:multi-wave}

J1652 belongs to the class of Extremely Red Quasars (ERQs), a population of luminous ($\ga 10^{47}$ \ergs) quasars at $z>2$ originally identified based on their extremely high infrared-to-optical ratios and peculiar optical spectra using the combination of data from Wide-field Infrared Survey Explorer (WISE; \citealt{wrig10}) and from the Baryon Oscillation Spectroscopic Survey (BOSS; \citealt{daws13}) of the 3rd generation of the Sloan Digital Sky Survey (SDSS-III; \citealt{eise11}). Specifically, ERQs are required to have $i-W3\ge 4.6$ mag (when both the optical $i-$band magnitude and the infrared $W3$ magnitude are on the AB system) and a rest equivalent width of CIV of $\ge 100$\AA. 

J1652 is a radio-intermediate \citep{hwan18} ERQ at z=2.9482. Initial near-infrared long slit spectroscopy revealed a very broad blueshifted component in the \oiii\ emission lines \citep{alex18}, with the overall width containing 80\% of line power of $w_{80}=1760$ \kms. This value is in the top 5\% compared to the luminous obscured quasars at $z<1$ \citep{zaka14}, but is comparable to those seen within the ERQ population, which presents the fastest \oiii\ kinematics of any quasar population \citep{zaka16b, perr19}. Rest-frame ultraviolet spectropolarimetric observations \citep{alex18} of J1652 with Keck reveal a 90$^{\circ}$ swing as a function of velocity in the polarization position angle for rest-frame UV emission lines (L$\alpha$, C IV and N V). The spectral shapes of the rest-frame ultraviolet lines and their polarization properties indicate that these emission features originate in a dusty equatorial outflow, moving at several thousand \kms\ on scales of up to a few tens of pc. The rest-frame UV continuum shows a strong polarization of 15\%, with an orientation of the scatterer relative to the nucleus, 220$^{\circ}$ East of North. 

The spectral shape of \oiii, with an extremely broad component blue-shifted relative to a narrower one, strongly suggests an outflow, however, the mass and the energetics of the outflow cannot be obtained without spatially resolved observations. We therefore obtained ground-based IFU observations \citep{Vayner21a} with the Gemini Near-Infrared Integral Field Spectrometer (NIFS) behind laser-guided adaptive optics, targeting rest-frame optical emission lines (\hb\ and \oiii) redshifted into the near-infrared ($K$-band). These observations revealed emission towards the south-west, extending out to $\sim 5$ kpc from the nucleus, consisting of both narrow and broad components. The narrow emission extends more towards the western direction, while the outflow extends almost entirely due south-west. \textit{Hubble Space Telescope} WFC3 IR observations tracing rest-frame B band reveal extended emission in the western direction likely originating from a tidal tail feature \citep{zaka19}. The B-band luminosity is above the median for the ERQs and is $\ga 10 L^*$ for galaxies at this redshift. \textit{Hubble Space Telescope} ACS observations through the F814W filter reveal extended emission in the south-west direction along PA 220$^{\circ}$ East of North, matching both the direction of the quasar-driven outflow and the scattering angle. The extended emission detected in ACS observations is likely entirely due to scattered light, confirming the geometric scenario where the outflow, and light scattering occur perpendicular to the central obscuration source. Based on X-ray observations from \textit{Chandra X-ray Observatory}, \textit{NuSTAR}, and \textit{XMM-Newton} J1652 is Compton thick \citep{goul18a,Ishikawa21} with a column density of 1$\times10^{24}~\rm cm^{-2}$ and X-ray luminosity $L_{2-10~keV}=1.27\times10^{45}$ \ergs. 

The first \textit{JWST} paper on J1652 \citep{Wylezalek22} focused on the discovery of a galaxy group around the quasar host galaxy that is likely a merger of two larger dark matter halos. Our second paper \citep{Vayner23b} focused detailed analysis of the system using the \qfit\ software \citep{ifsfit2014} focusing on ionized emission line ratios to decipher the source of ionization and warm ionized gas phase conditions across the quasar host galaxy and the neighboring galaxies. In the quasar host galaxy, we confirmed the existence of the ionization cone along the photon diffusion path from spectropolarimetric observations towards the southwest, in the same direction as the outflow. Orthogonal to the direction of the outflow, we discover kpc scale regions with emission line ratios consistent with radiative shocks, likely driven by the interaction of the outflow with the ISM. Towards the north-eastern direction, we discovered several star-forming clumps by isolating spaxels with emission line ratios indicative of ionization by massive young stars. We find a total dust-reddening corrected star formation rate (SFR) of 200 $M_{\odot}$ yr$^{-1}$. The ISM within these star-forming clumps shows relatively high electron densities, reaching up to 3,000 cm$^{-3}$. We find gas-phase metallicities ranging from half to third solar with a positive metallicity gradient in the direction away from the center of the star-forming region and V band extinctions up to 3 magnitudes. Neighboring galaxies in the path of the ionization cone show evidence of quasar photoionization in their ISM. In contrast, galaxies away from the path showcase a combination of ionization from AGN and stars. 

\pagebreak
\section{Analysis} \label{sec:analysis}

This section outlines our procedure to remove the bright quasar emission from the data and fit the underlying faint emission lines and stellar continuum from the host galaxy and the nearby companion galaxies and tidal tail features. 

\subsection{PSF subtraction}

To conduct PSF subtraction, we are utilizing the \qfit\ software package. The philosophy behind our PSF subtraction routine is that the inner few spaxels near the centroid of the quasar are almost entirely dominated by point source emission, allowing us to construct a ``quasar template" spectrum. We extract a spectrum using a 2-pixel radius centered on the quasar. We mask each bright emission line that we believe may be extended on galaxy scales using a 2000 \kms\ mask at the redshift of the quasar host galaxy. We create a signal-to-noise ratio map by dividing the peak of the \oiii\ 5007 \AA\ emission in each spaxel by the standard deviation at that channel. We then loop over all spaxels that show a signal-to-noise ratio $>3$ and fit the spectra by scaling the ``quasar template" with a series of multiplicative polynomials and exponential functions that account for variations in the spectral shape of point source emission in individual spaxels due to the spectral dependence of the NIRSpec PSF. The PSF subtraction makes use of the broad wings of Balmer emission lines and the bright quasar continuum as the anchors in fitting and subtracting the quasar template at each spaxel. 

We opt to use a 2-pixel radius to construct the quasar template in order to create a spectrum that is not heavily dominated by the spectral oscillating pattern produced by the under-sampling of the PSF by the native slicer plate scale of 100 mas. After extensive investigation, we find that the oscillating pattern is better mitigated by using the ``emsm" method in the ``Spec2Pipeline" to construct the data cubes. We subtract the PSF over the entire wavelength range from 1.66 - 3.17 \micron. \\

\qfit\ outputs a model of the NIRSpec IFU PSF. To better understand the quality of our PSF subtraction, we compared the model PSF data cube to that of the standard star observations over a similar wavelength range (1.66-1.9\micron) and find that azimuthally averaged profiles of the model PSF and the empirical PSF from the standard star agree on average to within 3\% in a 0.75 \arcsec\ radius (see Appendix \ref{sec:appendix_A}). More details on our PSF subtraction quality will be forthcoming in \qfit\ software paper.

\subsection{Emission line fitting}
We perform emission line fitting using the \qfit\ software, which subtracts the best-fit quasar template model and leaves behind emission from the spatially extended component. We fit all emission lines detected above $3\sigma$ simultaneously with Gaussian models. The velocity width and radial velocity offset for each line are fixed to the \oiii\ 5007 \AA\ line, selected due to its higher SNR at each spaxel. The model emission line spectrum is then convolved with the NIRSpec line-spread-function and sent into \texttt{lmfit} to be fit with the least-squares routine. We select the number of Gaussian components per emission line to minimize the final chi-squared value. The fluxes of all emission lines are set as free parameters except quantum mechanically locked lines such as \oiii 4959, 5007, and \nii\ 6548, 6583. We set a minimum flux ratio limit on the Balmer emission lines based on Case B recombination. For the \sii\ 6716, 6731 doublets, the flux ratio can vary between 0.4 and 1.5 governed by the gas-phase temperature and density conditions of the ISM. We also set a limit on the allowed range for the emission line velocity dispersion, between 40 \kms\ and 1500 \kms\ to avoid fitting noise spikes and potentially very broad non-astrophysical features caused by the oscillating pattern of the NIRSpec point-source spectra. 

In addition to fitting emission lines, we also fit a 3rd-order polynomial to account for any potential continuum emission from stellar light. We experimented with spaxel size to improve fitting fainter and more diffuse emission across the data cube but found that the best fitting results are obtained when using the 50 mas plate scale. This is due to both the complex emission line structure observed in the data cube with multiple emission line components per emission line and a large velocity  variation across the data cube, which causes artificial smearing of the line profiles when binning the data, making it more complicated for \qfit\ to fit the emission line profiles with multiple Gaussians.

\subsection{Emission line maps and search for multiple kinematic components}

After fitting the emission lines with Gaussian components, we create integrated intensity, radial velocity, and dispersion maps for each kinematic component for each fitted line. We fit a maximum of 3 Gaussian components to each emission line representing three distinct kinematic components. These three kinematic components can be identified by making specific radial velocity and dispersion cuts:

\begin{enumerate}
    \item We detect a high velocity offset ($|V_{50}|>$ 500 \kms) and generally narrow (V$_{\sigma}<$ 300 \kms) kinematic component. This kinematic component can then be further subdivided into three sub-components  at a velocity cut of 500 \kms $<$ V$_{50}<$ 700 \kms and 700 \kms $<$ V $<$ 900 \kms\ associated with four galaxies and their tidal tails due to interactions with J1652 quasar host galaxy.
    \item We detect a narrow (V$_{\sigma}<$ 300 \kms) and close to systemic velocity offset kinematic ($|V_{50}|<$ 500 \kms) component ascociated with the quasar host galaxy.
    \item We detect a broad (V$_{\sigma}>$ 300 \kms) kinematic component associated with the outflow in the host galaxy of J1652.
\end{enumerate}

\begin{figure*}
    \centering
    \includegraphics[width=1.0\textwidth]{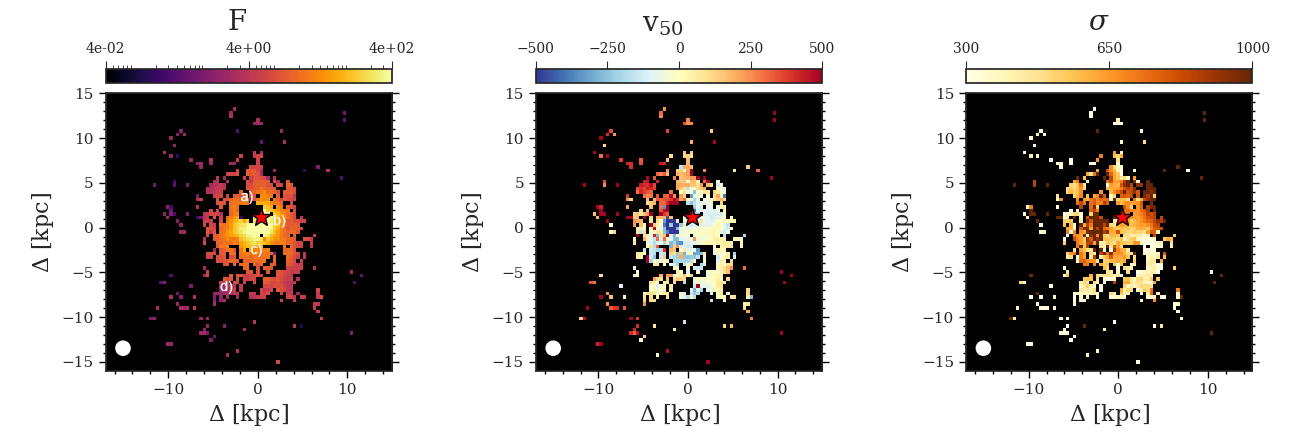}\\
    \caption{On the left, we present line-integrated \oiii\ 5007 \AA\ flux map in units of $\times10^{-17}$\surff, in the middle, we present the radial velocity centroid ($v_{50}$) map in units of \kms\ and on the right the velocity dispersion map for the kinematic components in J1652 associated with the broad extended emission. Regions for which spectra are extracted in Figure \ref{fig:outflow_spec} are labeled a-d. North is up, and east is to the left.}
    \label{fig:kinematic_component_maps}
\end{figure*}

\begin{figure}
	\centering
	\includegraphics[width=1.0\linewidth]{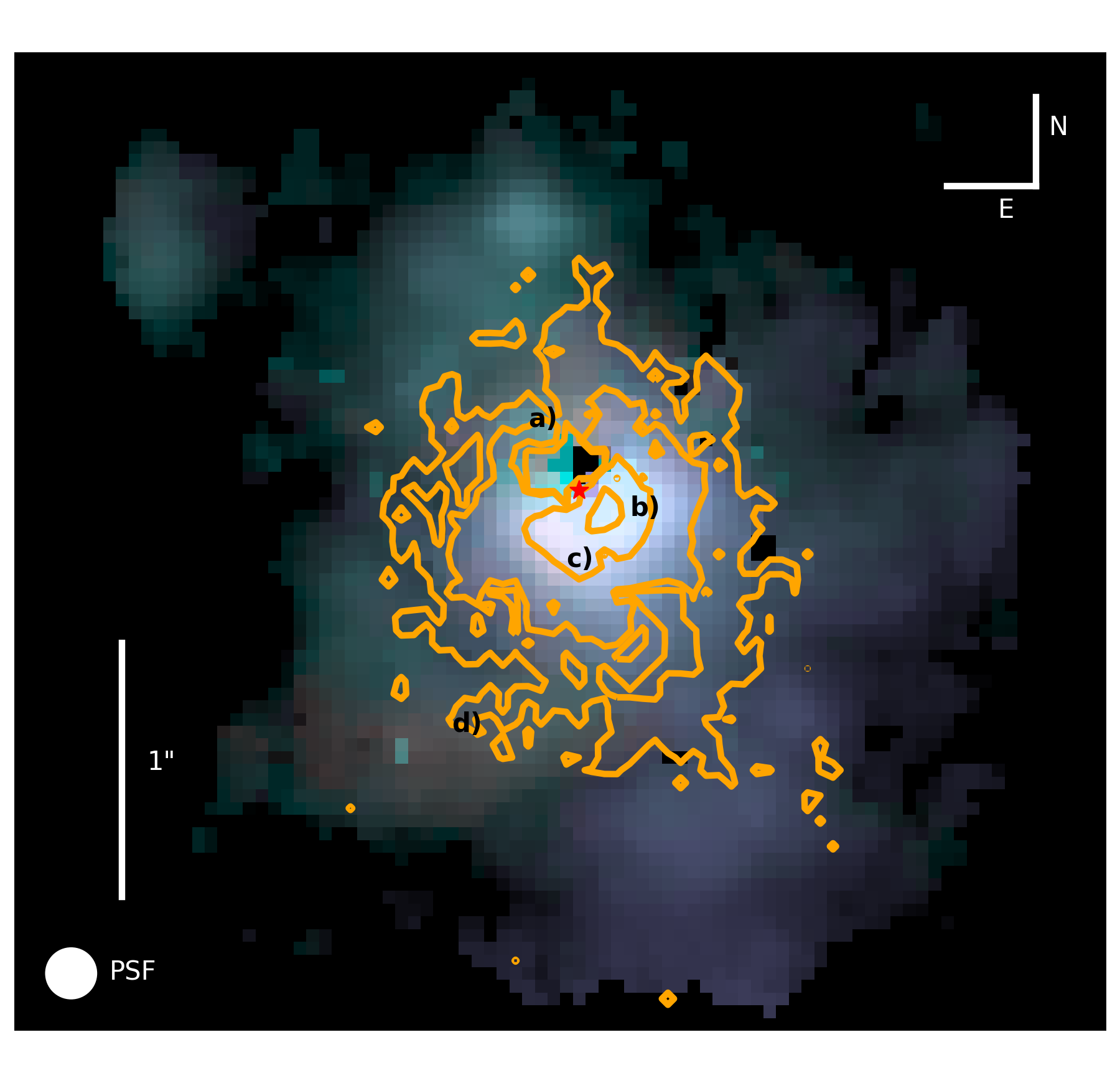}
	\caption{A color composite of total ionized gas emission integrated over all kinematic components in the J1652 quasar host galaxy and the immediate environment. \oiii\ is assigned to blue, \ha\ is assigned to green, \nii\ is assigned to red and \sii\ is assigned to orange. Orange contours represent the line integrated \oiii\ intensity associated with the extended broad emission from the galaxy scale outflow. Bar on the left represents 1 arcsecond. The FWHM of the PSF representing the effective resolution of our observations is shown in the lower left corner. Regions for which spectra are extracted in Figure \ref{fig:outflow_spec} are labeled a-d. North is up, and east is to the left.}
	\label{fig:threecolor}
\end{figure}

\begin{figure}
    \centering
    \includegraphics[width=1.0\linewidth]{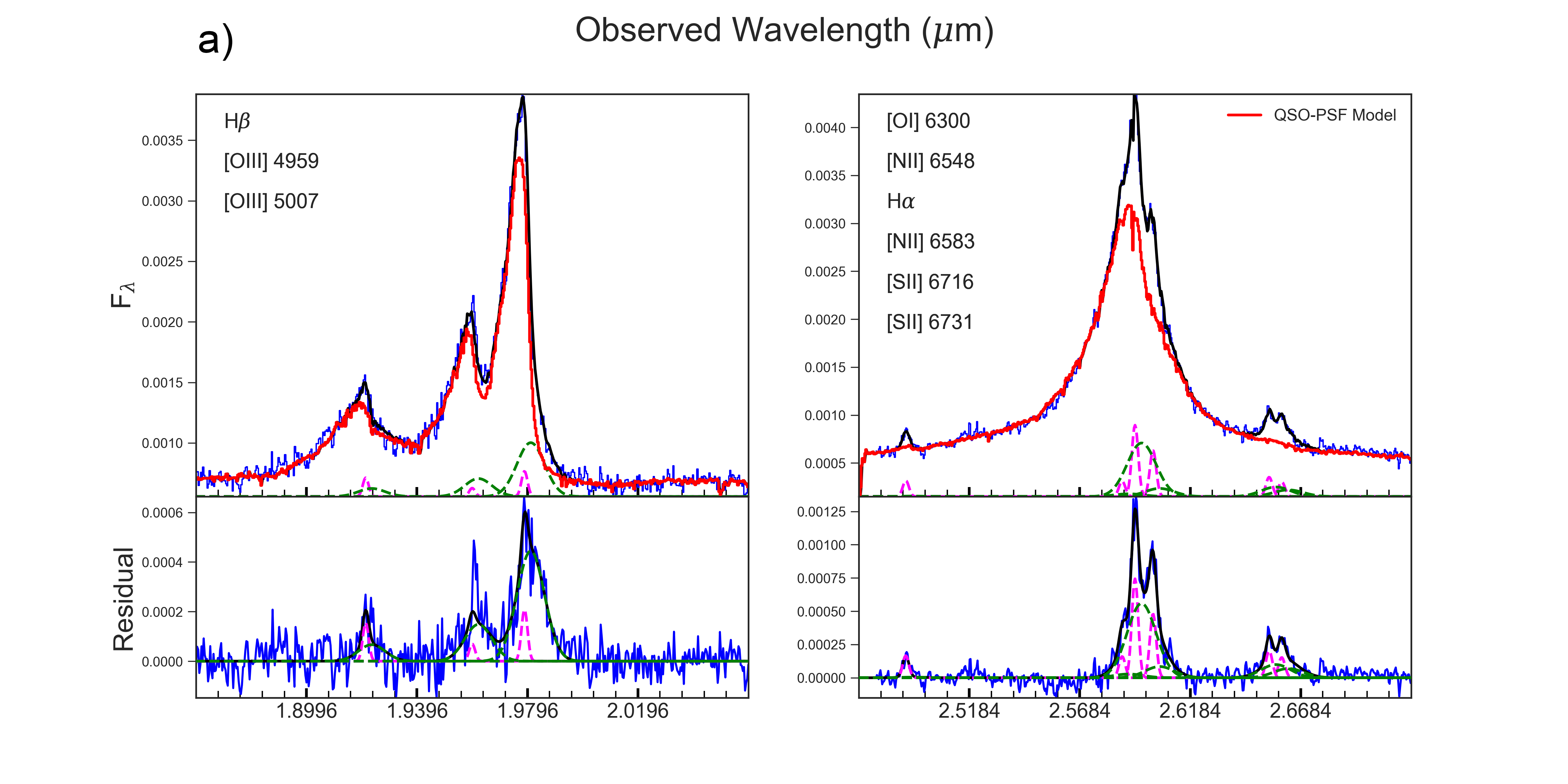}\\
    \includegraphics[width=1.0\linewidth]{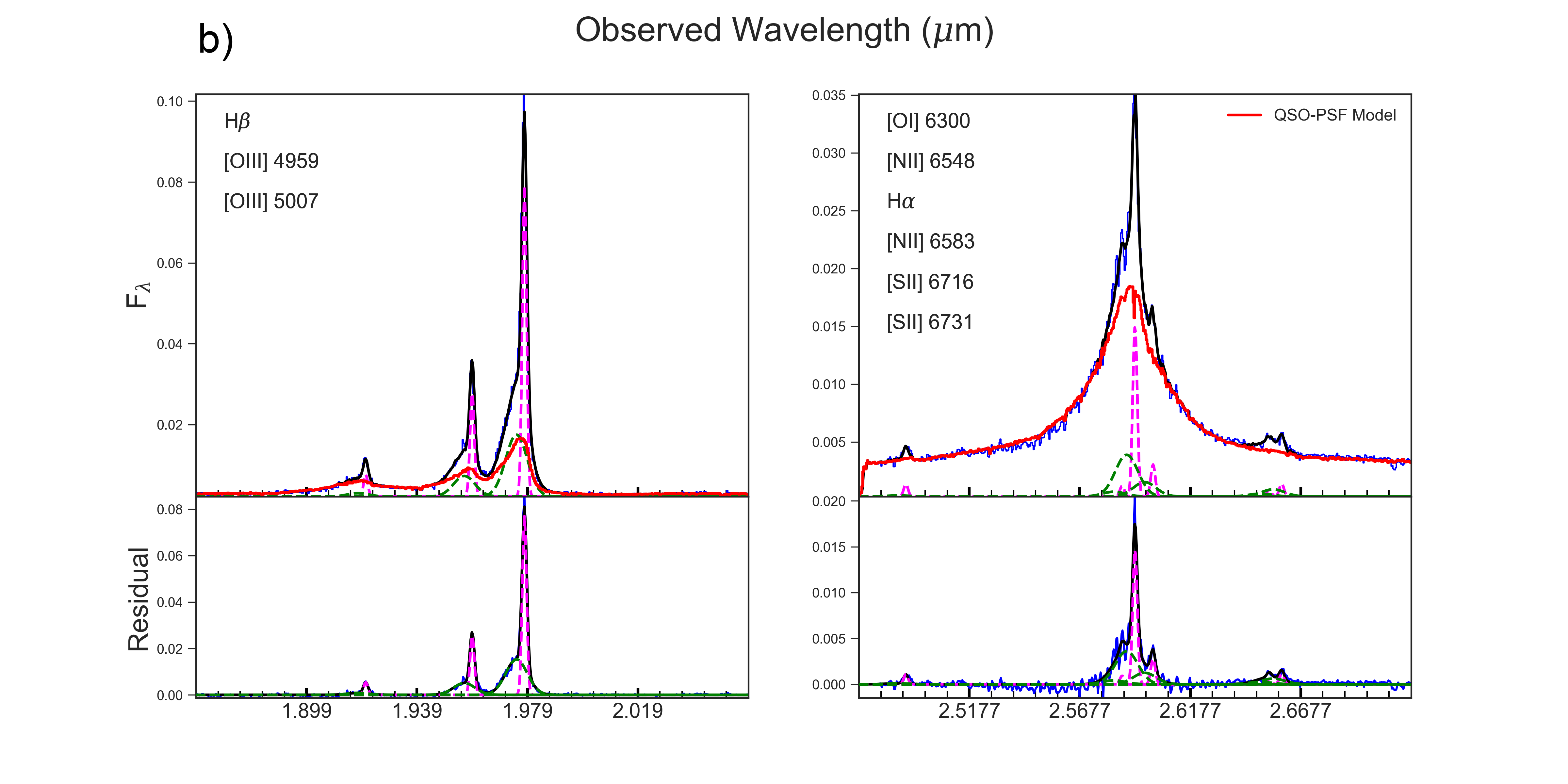}\\
    \includegraphics[width=1.0\linewidth]{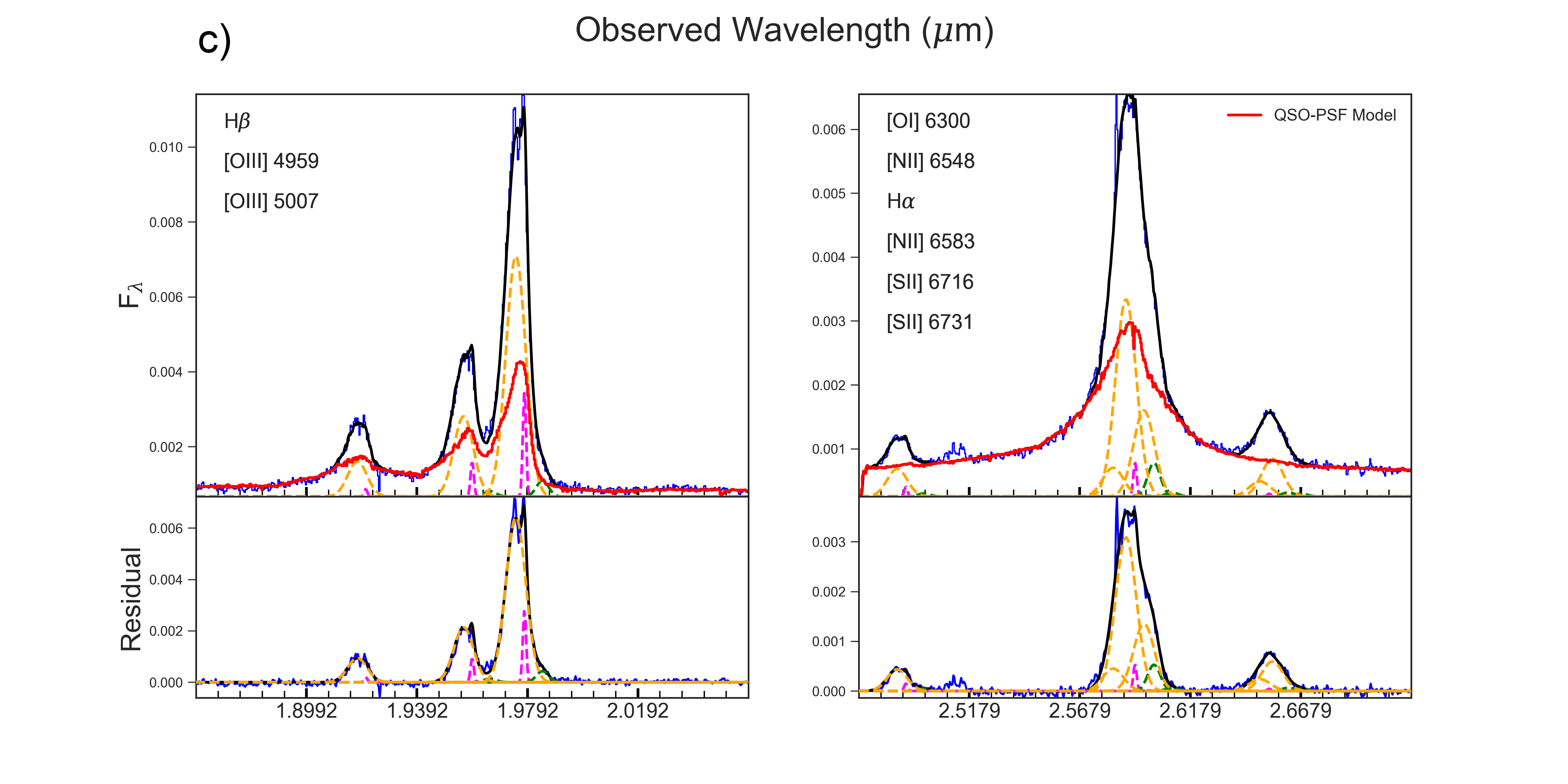}\\
    \includegraphics[width=1.0\linewidth]{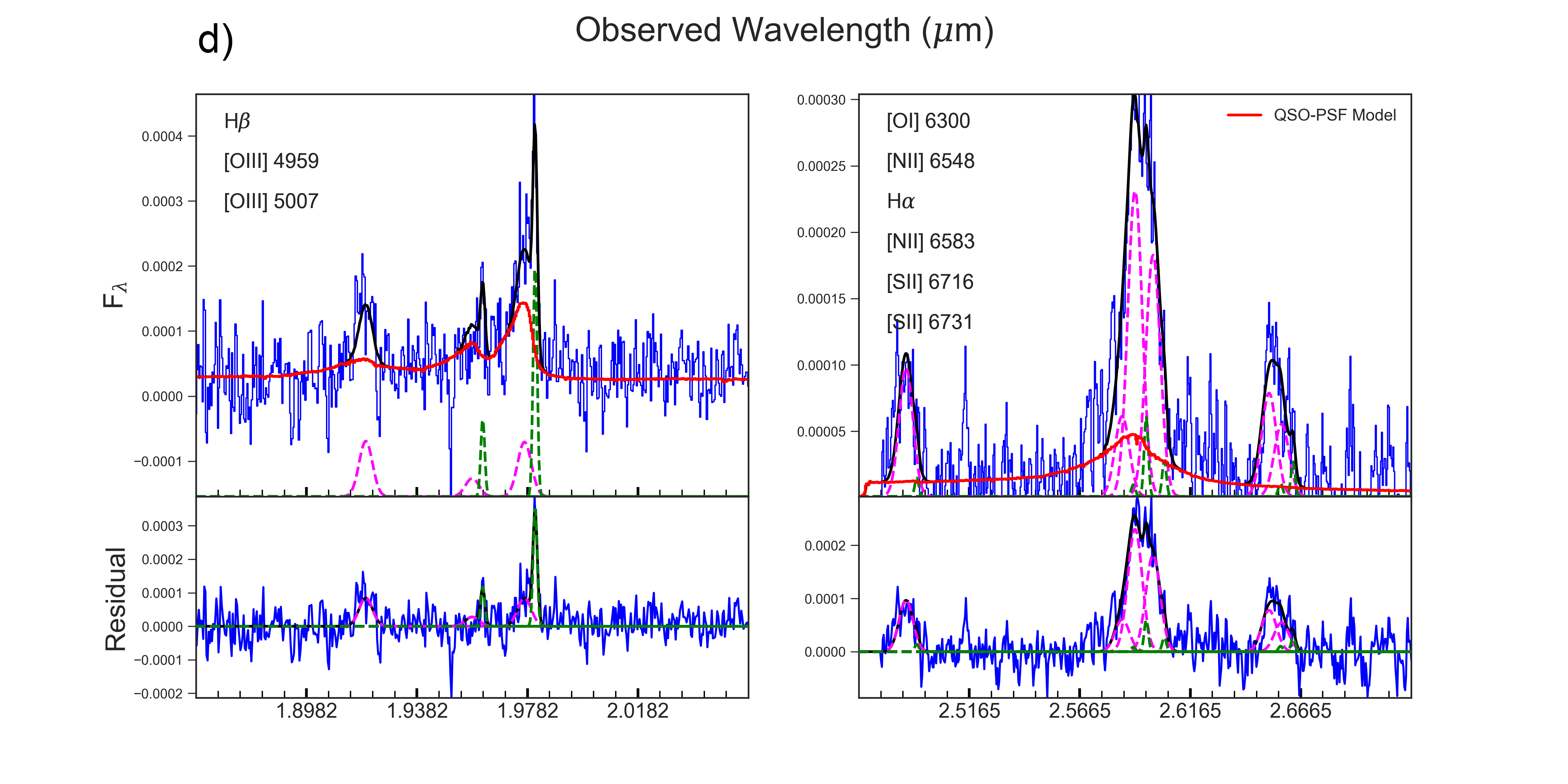}
    \caption{Example of spectra along with model fits across the outflow region in J1652 to the \hb, \oiii, \oi, \ha, \nii, and \sii\ emission lines. The top panel for each spectrum shows a multi-component Gaussian fit (dashed lines) to extended emission consisting of a narrow component for the quasar host galaxy or the merging system and a broader component associated with the galaxy scale outflow, the red curve shows the spatially unresolved emission fit to that spaxel part of the PSF subtraction routine in \qfit\ and the black curve shows the total fit. The bottom sub-panel shows the residual after subtracting the best-fit spatially unresolved emission. Similarly, the dashed curves show the individual Gaussian components fit to the spatially extended emission, while the black curve shows the total fit to the extended emission. Spaxel a) is part of the redshifted cone of the outflow region, spaxel b) is part of the maximum blueshifted region of the outflow, spaxel c) shows the intermediate velocity shift region while spaxel c) is near the southeast edge of the outflow.}
    \label{fig:outflow_spec}
\end{figure}

In Figure \ref{fig:kinematic_component_maps}, we present the integrated \oiii\ 5007 \AA\ intensity, radial velocity, and dispersion maps of the broad kinematic component part of the resolved extended outflow. In Figure \ref{fig:threecolor}, we present a color composite consisting of the \oiii, \ha, \nii, and \sii\ extended emission of the quasar host galaxy, the outflow, and the surrounding immediate quasar host galaxy environment. In Figure \ref{fig:outflow_spec}, we present the spectra along with their multi-component Gaussian fits in several spaxels in the quasar host galaxy outflow region to highlight the line morphology across the outflow. 

\subsection{Fitting the spatially unresolved nuclear emission:}
\label{sec:spatially_unresolved_emission}
In J1652, a significant amount of emission line flux is spatially unresolved at an angular resolution of 200 mas. In \citet{Vayner21a}, we discovered a large fraction of the broad emission associated with the outflow is found to be within the nuclear region ($<1$ kpc). In this section, we describe how we model and extract flux from the spatially unresolved component in this system.

For the spatially unresolved spectrum, we use the spectrum extracted from a radius of 2.5 pixels centered on the quasar used for PSF subtraction in Section \ref{sec:data-reduction}. We apply an aperture correction to this spectrum to calculate the total flux based on curve of growth analysis of the standard star over the same wavelength range. We use the \pyqsofit\ \citep{Shen19,Guo19} software to fit the quasar continuum using a combination of a power-law, Polynomial, and the Fe II emission. Channels free of broad and narrow emission lines from the quasar and the host galaxy are used for continuum fitting. We fit the broad line region Balmer emission lines, HeII 4687 \AA\ and HeI 5875 \AA\ emission lines, using a combination of two broad Gaussians. Intermediate velocity width lines arising from the outflow seen in \hb, \oiii, \ha\ and \nii\ are fit with a combination of two Gaussian lines, and their velocity width and offset are tied to the \oiii\ emission line. 

\begin{figure*}
    \centering
    \includegraphics[width=1.0\linewidth]{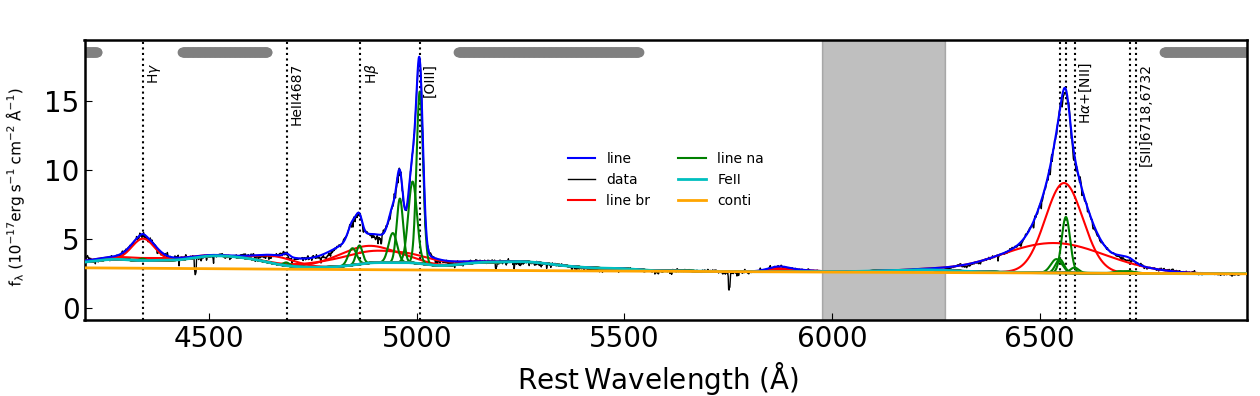}
    \caption{Spatially unresolved spectrum of J1652 along with the best continuum and emission line fit using the \pyqsofit\ software. The spectrum is shown in black, the best fit total spectrum is shown in blue. The best fit quasar powerlaw continuum is shown in orange, the Fe II emission line template fit is shown in teal. Broad emission line arising from the quasar broad line region are shown in red, while the intermediate velocity component associated with the outflow are shown in green. Several key emission lines are labeled. Regions used for continuum fitting are highlighted with the gray bars. Wavelengths with no spectral coverage due to the NIRSpec detector gap are shown in gray.}
    \label{fig:unresolved_spec}
\end{figure*}

\section{Measuring outflow rates and energetics}\label{sec:out-energy}

We detect broad $V_{\sigma}>300$ \kms\ emission across a large fraction of the J1652 host galaxy. We measure a maximum extent of 10 kpc towards the south-east direction in the same general path as the ionization cone, however, extending nearly 2 times further than what was previously measured in the ground-based NIFS observations \citep{Vayner21a}. The outflow component generally shows a consistent broad emission, with a slight decrease in the velocity dispersion as a function of radius towards the southwest. In the section below, we outline how we estimate the mass, outflow rate, and energetics of the ionized outflow, compare these values to theoretical models and discuss the potential mechanisms responsible for driving the galaxy scale outflow. 

We measure the ionized gas mass using the \ha\ emission lines, assuming recombination as the primary source of line production. For \ha, we assume case B recombination and follow the methodology of \cite{oste06}, assuming that each emitting cloud within a given aperture has the same electron density and that the electron density is constant across each aperture. We further assume a solar abundance for helium and that the gas in the outflow region is fully ionized, with helium being an equal mix of HeII and HeIII. Under these assumptions, we get the following relationships:

\begin{equation}\label{equation:ionized_gas_mass}
    M_{\alpha,\delta}^{Ionized} = 1.4 \bigg(\frac{m_{p}L_{H\alpha_{\alpha,\delta}}}{j_{H_{\alpha}}n_{e}}\bigg)
\end{equation}

\noindent where $\rm L_{H\alpha_{\alpha,\delta}}$ and $\rm n_{e}$ are \ha\ luminosity per spaxel and the average electron density over the outflow region measured in Paper 1 \citep{Vayner23b}. We assume a range of (1-2)$\times10^{4}$ K for the electron temperature, which constrains the \ha\ line emissivity to (1.8-3.53)$\times10^{-25}\rm~erg~cm^{3}~s^{-1}$ for an electron density of $\sim10^{2-3}$ \eden. We compute the emissivities using the \texttt{PyNeb} package \citep{Luridiana15}.\\

We compute the total outflow rate by integrating all the outflow spaxels using the following equation:

\begin{equation}\label{equation:outflow-simple}
    \dot{M}^{Ionized}_{Total}=\sum{ \frac{M_{\alpha,\delta}v_{\alpha,\delta}}{R_{\alpha,\delta}}},
\end{equation}

\noindent where $M_{\alpha,\delta}$ is ionized gas mass per spaxel and $v_{\alpha,\delta}$  is the outflow velocity in each individual spaxel computed using $v_{\alpha,\delta} = v_{r} + \sigma_{r}$ where $v_{r}$ is the radial velocity measured using the centroid of the Gaussian fit and $\sigma_{r}$ is the radial velocity dispersion (Figure \ref{fig:kinematic_component_maps}). Our reasoning for calculating the outflow velocity in this manner is to account for possible inclination effects. Likely, the lower velocity structure is caused by the projection of the conical outflow along our line of sight. The wings of the emission line profile likely encode the true outflowing velocity of the ionized gas \citep{cano12,gree12}. We use the combination of line centroid and dispersion to approximate the maximum velocity achieved in the wings of the emission line \citep{rupk13a}. For the radius ($R_{\alpha,\delta}$), we use the distance from the quasar to the individual spaxel. For the spatially unresolved component of the outflow, we use the PSF FWHM/2 value for the radius. Additionally, we measure the instantaneous outflow rate based on the time scale it takes for the total mass per spaxel to cross the distance of a single spaxel:

\begin{equation}\label{equation:outflow-simple}
    \dot{M}_{\alpha,\delta}=\frac{M_{\alpha,\delta}v_{\alpha,\delta}}{\Delta R},
\end{equation}

\noindent For the radius $\Delta R$ we use the physical size of each individual spaxel. We then compute the integrated and instantaneous momentum flux and kinetic luminosities of the outflow using the following equations:

\begin{equation}\label{mom}
    \dot{P}_{\alpha,\delta}^{outflow}=\dot{M_{\alpha,\delta}}\times v_{\alpha,\delta}
\end{equation}

\begin{equation}\label{en}
    L_{\alpha,\delta}^{kinetic} = \frac{1}{2} \dot{M_{\alpha,\delta}}\times v_{\alpha,\delta}^{2}.
\end{equation}

\begin{figure*}
    \centering
    \includegraphics[width=1.0\textwidth]{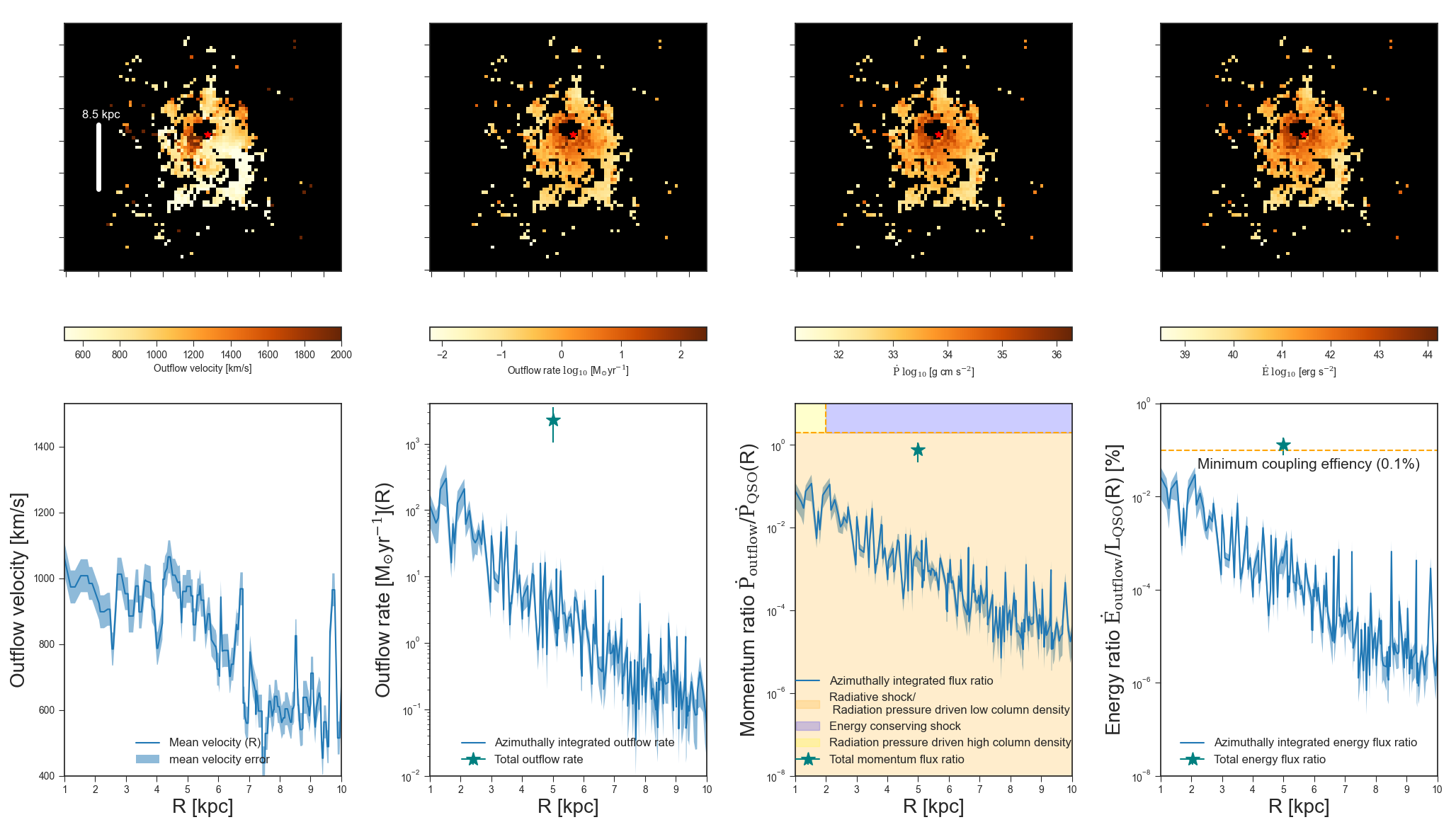}
    \caption{Spatially resolved energetics of the ionized outflow in the J1652 quasar host galaxy. The top row shows the outflow velocity and maps of the instantaneous outflow rate, momentum flux, and kinetic luminosity per spaxel. The bottom panel shows the median outflow velocity, the azimuthally integrated outflow rate, momentum flux ratio, and energy flux ratio profiles as a function of radius measured in concentric shells. The stars in the lower panel show the respective integrated quantities over the entire outflow. Colors on the momentum flux ratio map indicate regions of expected momentum flux ratios based on different driving mechanisms behind the galaxy scale outflow based on the total spatially integrated energetics. The energy flux ratio diagram in the lower right panel shows the minimum coupling efficiency required between the quasar bolometric luminosity and the quasar-driven outflow for there to be an impact on the gas in the host galaxy based on the consensus of theoretical works.}
    \label{fig:resolved_energetics}
\end{figure*}

\noindent We present per spaxel values for the instantaneous energetics rates and take a sum to present the total values. The top panel in Figure \ref{fig:resolved_energetics}, shows the outflow velocity, outflow rate, momentum flux, and kinetic luminosity per spaxel. Integrated outflow rate and energetics are presented in Table \ref{tab:outflow-prop} and include the values from the spatially unresolved component discussed in Section \ref{sec:spatially_unresolved_emission}. For comparison with simulations, we also measure the outflow rate, momentum flux ratio, and kinetic luminosity ratio profiles by measuring the instantaneous energetics in concentric shells as a function of the radius from the quasar: 

\begin{equation}\label{equation:outflow-shell}
    \dot{M}(R)=\sum{\frac{M_{\alpha,\delta}(R)v_{\alpha,\delta}(R)}{\Delta R}},
\end{equation}

\begin{equation}\label{equation:mom-shell}
    \dot{P}(R)/L_{bol}/c=\sum{\frac{M_{\alpha,\delta}(R)v_{\alpha,\delta}^{2}(R)}{\Delta R}}/L_{bol}/c,
\end{equation}

\begin{equation}\label{equation:energy-shell}
    {L(R)}/L_{qso}=\frac{1}{2}\sum{\frac{M_{\alpha,\delta}(R)v_{\alpha,\delta}^{3}(R)}{\Delta R}}/L_{bol}
\end{equation}

\noindent The momentum and energy flux ratios are measured relative to the photon momentum flux of the quasar ($L_{bol}/c$), while the kinetic luminosity ratio is measured relative to the bolometric luminosity ($L_{bol}$) of the quasar in units of percentage. We assume that the instantaneous energetics are measured at a single time stamp at the dynamical time scale of the outflow ($t_{outflow} = R_{outflow}/v_{outflow} \sim 5$Myr). We present the outflow velocity profile, outflow rate, momentum, and kinetic luminosity ratios in the bottom panel of Figure \ref{fig:resolved_energetics}. We also present the integrated values, labeled as stars in the bottom row of Figure \ref{fig:resolved_energetics}, measured at the dynamical radius of the outflow. We generally see a trend of decreasing instantaneous outflow rate, velocity, and energetics as a function of the radius from the quasar.

\begin{deluxetable*}{llllllll}

\tablecaption{Integrated outflow properties of J1652 \label{tab:outflow-prop}. R$\rm_{out}$ is the maximum observed radial extent of the outflow. $V_{out}$ is the average outflow velocity. $dM/dt\rm_{H\alpha}$ is the outflow rate derived from the \ha\ line. $\dot{P}_{outflow}$ is the momentum flux of the outflow and $\dot{E}_{outflow}$ is the kinetic luminosity of the outflow. L$\rm_{AGN}$ is the bolometric luminosity of the quasar. $\frac{\dot{E}_{outflow}}{\dot{L}_{AGN}}$ is the energy coupling efficiency between the kinetic luminosity of the outflow and bolometric luminosity of the quasar. $\frac{\dot{P}_{outflow}}{\dot{P}_{AGN}}$ is the momentum flux ratio between the ionized outflow and the photon momentum flux from the quasar accretion disk.}

\tablehead{\colhead{$R\rm_{out}$}&
\colhead{$V\rm_{out}$}&
\colhead{$dM/dt\rm_{H\alpha}$}&
\colhead{$\dot{P}_{outflow}$}&
\colhead{$\dot{E}_{outflow}$}&
\colhead{$L_{AGN}$}&
\colhead{$\frac{\dot{E}_{outflow}}{\dot{L}_{AGN}}$}&
\colhead{$\frac{\dot{P}_{outflow}}{\dot{P}_{AGN}}$}\\
\colhead{kpc}&
\colhead{\kms}&
\colhead{\myr}&
\colhead{$10^{36}$dyne}&
\colhead{$10^{43}$ \ergs}&
\colhead{$10^{47}$ \ergs}&
\colhead{$\%$}&
\colhead{}}
\startdata
10 & 790$\pm$10  & $2300\pm$1200 &  14$\pm$7 & 70$\pm$30 & 5.37$\pm$0.1 & 0.1$\pm$0.05 & 0.8$\pm$0.4\\
\enddata
\end{deluxetable*}

\section{Discussion} \label{sec:discussion}

\subsection{Star formation as potential driving mechanism behind the galaxy scale outflow:}
Several driving mechanisms can cause the galaxy scale outflow in the host galaxy of J1652. Radiation pressure from massive O-type stars and supernovae explosions are typically the two dominant sources of galaxy scale outflows due to star formation processes. Based on Starburst99 models \citep{Leitherer99}, radiation pressure from massive young stars can inject the following amount of momentum flux into the surrounding ISM:
\begin{equation}
    \dot{P}_{SFR} = 1.5\times10^{33}\frac{\dot{M}_{SFR}}{1M_{\odot}yr^{-1}} \rm dyne/s.\label{eq:pdot_SFR_SFR}
\end{equation}

\noindent assuming a Kroupa initial mass function \citep{kroup01}, and writing the equation in terms of the star formation rate ($\dot{M}_{SFR}$). Detailed hydrodynamical simulations find that if the gas is optically thick to infrared photons, scattering off dust grains can increase the injected momentum flux by up to two times on kpc scales \citep{thom15,Costa18}. While at early times (4-6 Myr) in the star formation process, radiation pressure from massive stars may dominate the momentum flux injection rate, at later times, supernova explosions will be the dominant process. According to detailed numerical simulations, \cite{Martizzi15,kim15} both find a momentum flux injection rate of 

\begin{equation}
    \dot{P}_{SNe} = 7.5\times 10^{33} \frac{\dot{M}_{SFR}}{1M_{\odot}yr^{-1}} \left(\frac{n}{100\,{\rm cm}^{3}}\right)^{-0.18}
    \left(\frac{Z}{Z_\odot}\right)
    \rm dyne/s \label{eq:SNe_mom}.
\end{equation} 

\noindent assuming one supernova per 100 years with an SFR of 1 \myr \citep{Vayner21c}.  By isolating individual spaxels that show emission line ratios consistent with photoionization by stars, we find a total SFR of 200 \myr, which would inject a total momentum flux of 1.5$\times10^{36}$ dyne/s both from radiation pressure and supernovae explosions. The total momentum flux of the ionized outflow is ($14\pm7)\times10^{36}$ dyne/s, which is higher than the momentum flux rate that can be provided through stellar feedback. Furthermore, the outflow is clearly emanating from the location of the quasar, and the star-forming region is found to be offset from the quasar by 6 kpc to the northeast. Hence, using both the spatial information of the star-forming region, the geometry of the outflow, and the measured energy injection rate indicates that the outflow is more likely to be driven by the quasar. We find no strong evidence for star formation in the vicinity of the quasar; all of the emission in the central few kpc is predominantly ionized by the quasar. However, we cannot exclude a powerful nuclear starburst that can potentially drive the observed outflow. A nuclear-obscured star-forming region would require a star formation rate $>2000$\myr\ to drive the current ionized outflow.

\subsection{Quasar as the driving mechanism behind the galaxy scale outflow:}

With a bolometric luminosity of 5.4$\times10^{47}$\ergs and a photon momentum flux of 1.8$\times10^{37}$dyne/s the quasar has both enough energy and momentum to drive the observed galaxy scale outflow. The quasar can drive the outflow through several different mechanisms. To differentiate between them we look at the total amount of momentum flux and kinetic luminosity that they can deposit into the galaxy scale outflow at different times in the lifetime of the outflow and how the energetics are distributed through the outflow at any given moment. 

Outflow driven through radiation pressure can deposit as much as 2$\times$\momfluxagn on kpc scales through trapping of infrared photons in very high column density environments \citep{thom15,Costa18}. Generally, the total \momfluxratio peaks at around 2 at early times in the lifetime of the outflow, when the majority of the outflow material is on 1-2 kpc scales \citep{thom15,Costa18}. The \momfluxratio\ profile of the outflow at any given time tends to have a local maximum at smaller radii with a negative slope towards larger scales. 

Radiation pressure driven winds from the quasar accretion disk or jets can shock the immediate ISM and drive a powerful shock that can sweep up material and cause the observed galaxy scale outflow. The nuclear density, wind/jet velocity, and the velocity of the resulting shock govern whether the resulting shock expands adiabatically and cools rapidly and efficiently or doesn't cool efficiently \citep{fauc12b,Costa14} and expands in an energy conserving manner. Both resulting shocks can drive galaxy-scale outflows, however, the energy-conserving shock tends to drive more powerful galaxy-scale outflows as they retain a large fraction of the initial energy provided by the driving mechanism. The total \momfluxratio\ for energy conserving outflow ranges from 7-10 when the outflow reaches kpc scales \citep{fauc12b,Costa14}, with a \momfluxratio\ profile of the outflow at any given time tends to have a positive slope towards larger radii. 

Given our integrated \momfluxratio\ of $\sim1$ and an observed trend where \momfluxratio\ decreases with radius from the quasar, if the ionized gas phase of the outflow contains the largest fraction of the energetics then the outflow had to be driven through radiation pressure by trapping infrared photons in a relatively high $\sim 10^{23}$cm$^{-2}$ column density environment on small ($<0.2$kpc) spatial scales \citep{thom15}. If we assume that our median electron density is not a true representative of the electron density over the outflow region and the outflow is driven by an energy-conserving shock, then the electron density needs to be a factor of 5 lower on $1$ kpc scales and a factor of 100-10,000 lower on scales $>1$ kpc. This is contradictory with our measurements of the electron density in parts of the quasar host galaxy in the vicinity of the outflow and in spaxels where the velocity dispersion in the outflow component is narrow enough to enable us to measure the electron density using the \sii\ emission line ratio. The median electron density that we measure is consistent with what has been found in other outflows at high redshifts \citep{Vayner17,Forster19}, with a general consensus that the electron density in the inner few kpc can get as high as 1,000 \eden. However, there are several notable cases of well spatially resolved outflows in nearby galaxies, such as M82 \citep{heck90}, where it is clear that the electron density in the outflow drops with radius. 

Very likely, our estimated outflow energetics in the ionized gas phase are somewhat underestimated, likely because even with \textit{JWST} we are biased to detecting the more dense and more luminous recombination radiation in the outflow. A possible scenario is that if an energy-conserving shock drives the outflow, a large fraction of the momentum flux of the outflow needs to be in either a cold molecular, neutral, or highly ionized hot gas phase. Indeed recent observations studying molecular outflow at both low and high redshift, find a substantial fraction of the outflow energetics to be in the colder temperature gas phases \citep{Vayner17,Brusa18,Herrera-Camus19,Vayner21mol}, consistent with theoretical work of multi-phase quasar-driven outflow \citep{Richings18,Richings20}. Likely, just observing the warm ionized gas phase is not giving us a true picture of the total energetics of the outflow in J1652. Future ALMA observations of molecular gas traced through rotational lines of carbon monoxide (CO) are necessary to trace the cold molecular \htwo\ gas phase in the outflow, which may contain a substantial fraction of the energetics, providing us with a clearer picture of the driving mechanism in the J1652 quasar host galaxy.

\subsection{Quasar driven outflow coupling efficiency}
Nevertheless, we find a coupling efficiency between the kinetic luminosity of the quasar-driven outflow and bolometric luminosity of the quasar of $0.1\%$, which is at the minimum value required in simulations for outflow to have a substantial impact on the star formation in their host galaxies and help establish the local scaling relation between the mass of the supermassive black hole and the mass and velocity dispersion of the bulge \citep{choi12,hopk12}. In \citet{Vayner23b}, we detected enhanced \siiha\ and \oiha\ emission line ratios at larger velocity dispersions in the quasar host galaxy perpendicular to the direction of the outflow. The line ratios are also consistent with models of radiative shocks in the ISM and we interpret these results as the galaxy scale outflow causing shocks in the ISM at a wider angle outside the ionization cone of the quasar and ahead of the outflow. The enhanced velocity dispersion in these regions likely indicates additional turbulence in the ISM. We can estimate the kinetic energy of the shocked region orthogonal to the outflow. Using the line ratio and velocity dispersion we isolate the shocked region in \citet{Vayner23b} and measure an \ha\ line flux of 2.7$\times10^{-17}$ \ferg\ which translates to a line luminosity of $2\times10^{42}$ \ergs. We can measure the total gas mass in the shocked region using Equation \ref{equation:ionized_gas_mass} and total kinetic energy due to turbulence using the following equation:

\begin{equation}
    E_{turb} = 3/2 M_{ionized}\times\sigma_{m}^{2}
\end{equation}

\noindent where $M_{ionized}$ is the mass of the ionized gas over the shocked region, and $\sigma_{m}$ is the averaged velocity dispersion measured over the shocked regions of 230\kms. The factor of 3 accounts for the fact that we measure only the radial component of the velocity dispersion. We are likely overestimating the velocity dispersion since our measured value also contains thermal velocity dispersion and velocity dispersion due to gravitational motion. However, both of these velocity dispersions are expected to be far lower than the combined value. Our estimate on mass is a lower limit since our observations only allow us to account for the warm ionized component. We measure a total kinetic energy of 1.7$\times10^{56}$ ergs, while the kinetic energy of the quasar-driven galaxy scale outflow is 7$\times10^{57}$ ergs, indicating that the outflow has sufficient energy to drive the observed shock and enhanced turbulence in the ISM in the direction perpendicular to the outflow. This indicates that the outflow can drive shocks and turbulence that prolong the time for gas to cool efficiently, directly showcasing the impact of quasar-driven outflow in an extremely red quasar on the ISM properties of the quasar host galaxy.

\section{Conclusions} \label{sec:conclusions}

In this paper, we present {\it JWST} NIRSpec IFU observations of J1652, a powerful extremely red quasar at redshift at $z=2.94$. J1652 belongs to the population of extremely red quasars, which are selected from optical and infrared surveys based on their colors and rest-frame ultraviolet emission line properties \citep{ross15, hama17}. Extremely red quasars contain some of the fastest moving outflows of any quasar sample to date \citep{perr19} among the obscured quasar population. In this paper, we map the location of the fast-moving outflow across the quasar host galaxy using optical emission lines redshifted into the near-infrared at kpc scale resolution and we find:

\begin{enumerate}
    \item We detect the ionized outflow out to 10 kpc away from the quasar with outflow velocities ranging from 500-1400 \kms\ with a generally decreasing trend as a function of projected radial distance from the quasar. 
    \item We measure average instantaneous outflow rates of 0.1-100 \myr\ per spaxel and a total integrated outflow rate of 2300 \myr.
    \item We do not find sufficient energy and momentum injection rates from supernova-driven feedback to explain the observed galaxy scale outflow; hence the quasar is the likely primary driving source.
    \item A momentum flux ratio between the ionized outflow and the quasar accretion disk of 0.8 signals that the outflow is most likely driven by quasar radiation pressure on dust grains in a relatively high column density environment. If the outflow is driven by an energy-conserving adiabatic shock caused by either a jet or a nuclear disk wind, then a substantial amount ($>5$\momfluxagn) is required to be in other gas phases in the outflow. Nevertheless, our estimated energetics based on the ionized gas phase are likely lower limits.
    \item We measure a coupling efficiency between the quasar bolometric luminosity and the outflow of 0.1\%, which, based on simulations of quasar feedback, should be sufficient to cause negative quasar feedback in the host galaxy.
    \item Finally, we find that the galaxy scale quasar-driven outflow has sufficient energy to drive the shocks observed perpendicular and ahead of the outflow in the quasar host galaxy, signaling evidence for the direct impact of the outflow on the ISM; driving turbulence and prolonging the time it takes for the gas to cool efficiently.
\end{enumerate}

\begin{acknowledgments}
A.V., N.L.Z., Y.I., S.S. and N.D. are supported in part by NASA through STScI grant JWST-ERS-01335. N.L.Z further acknowledges support by the Institute for Advanced Study through J. Robbert Oppenheimer Visiting Professorship and the Bershadsky Fund.
\end{acknowledgments}

\vspace{5mm}
\facilities{JWST(NIRSpec), HST(WFC3) }\\
The data is available at MAST: \dataset[10.17909/qacq-9285]{\doi{10.17909/qacq-9285}}
\software{astropy \citep{Astropy2013, Astropy2018},  reproject \citep{thomas_robitaille_2023_7584411}, \qfit \citep{ifsfit2014}, \pyqsofit \citep{Shen19,Guo19}}

\pagebreak
\appendix

\section{PSF radial profiles:}
\label{sec:appendix_A}
\begin{figure}
    \centering
    \includegraphics[width=0.5\textwidth]{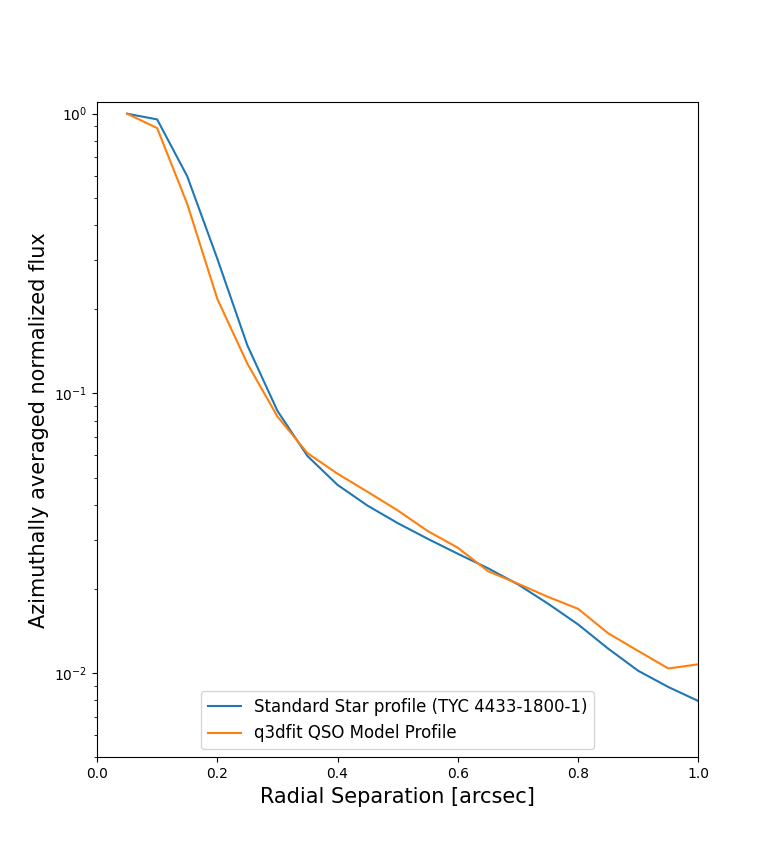}
    \caption{A comparison between the empirical and reconstructed \qfit\ J1652 PSF radial profile derived by integrating both data cubes in the same wavelength range. Each radial profile is normalized to the peak flux.}
    \label{fig:psf_comp}
\end{figure}

\begin{figure}
    \centering
    \includegraphics[width=0.5\textwidth]{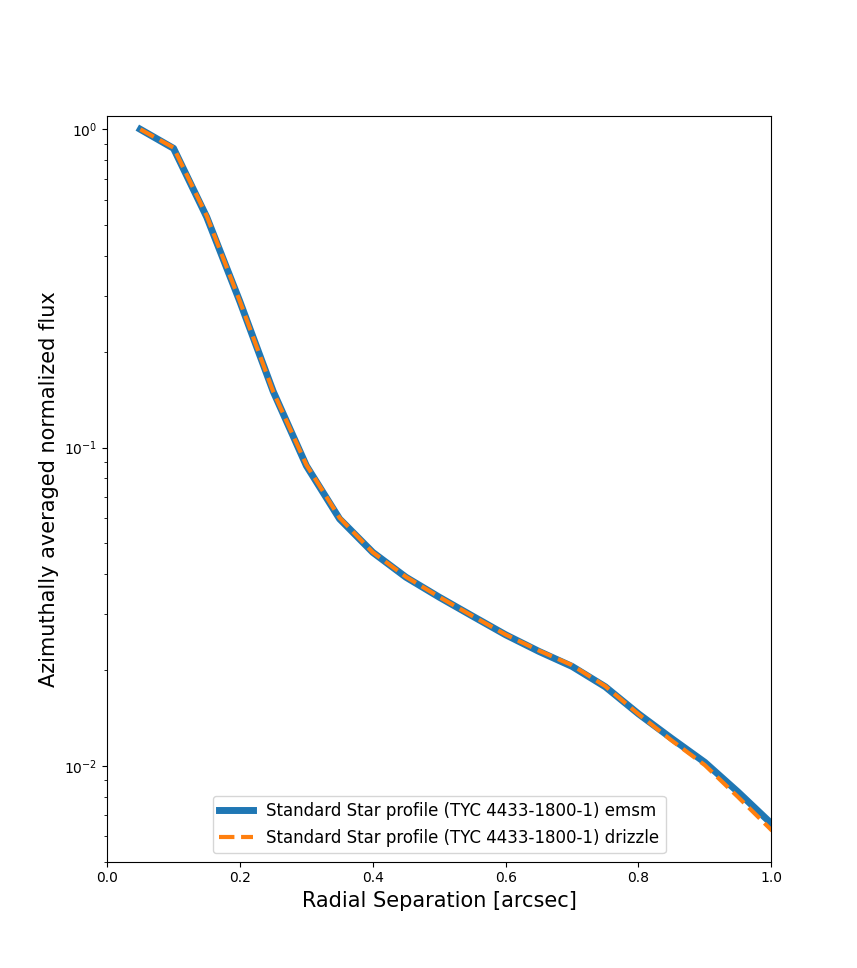}
    \caption{A comparison between the radial profiles of the standard star with the cube constructed using the ``drizzle" vs. ``emsm" method in ``Spec2Pipeline". Each radial profile is normalized to the peak flux.}
    \label{fig:emsm_vs_drizzle}
\end{figure}

We compare the radial profile of the PSF reconstructed using \qfit\ to the empirical PSF from standard star observations of TYC 4433-1800-1. The reconstructed and empirical PSF images are created at the same wavelength range by integrating each data cube along the wavelength axis. Both radial profiles are constructed by averaging the flux in azimuthal annuli with a width of 0.05\arcsec\ centered on each PSF. We present these results in Figure \ref{fig:psf_comp}. Similarly, we create radial profiles for the standard star where the data cubes were constructed using both the ``drizzle" and the ``emsm" method in ``Spec2Pipeline" and present the results in Figure \ref{fig:emsm_vs_drizzle}. Both profiles show an excellent agreement with the ``emsm" constructed data cube showing a slightly broader PSF.

\restartappendixnumbering

\bibliography{bib}{}
\bibliographystyle{aasjournal}

\end{document}